\documentclass[9pt]{article}

\usepackage[utf8]{inputenc}
\usepackage{amsmath,amsfonts,amssymb}
\usepackage{graphicx}
\usepackage[colorlinks=true, allcolors=blue]{hyperref}
\usepackage{authblk}
\usepackage[sort&compress,numbers]{natbib}
\newcommand{\ie}{\textit{i.e.}, }

\title{Optimal compressive multiphoton imaging at depth using single-pixel detection}

\author[1]{Philip Wijesinghe}               
\author[1]{Adri\`{a} Escobet-Montalb\'{a}n} 
\author[1]{Mingzhou Chen}                   
\author[2]{Peter R T Munro}                 
\author[1]{Kishan Dholakia}                 
\affil[1]{SUPA, School of Physics and Astronomy, University of St Andrews, North Haugh, St Andrews, KY16 9SS, UK}
\affil[2]{Department of Medical Physics and Biomedical Engineering, University College London, Gower Street, London WC1E 6BT, UK}

\begin{document}
\maketitle

\begin{abstract}

Compressive sensing can overcome the Nyquist criterion and record images with a fraction of the usual number of measurements required. However, conventional measurement bases are susceptible to diffraction and scattering, prevalent in high-resolution microscopy. Here, we explore the random Morlet basis as an optimal set for compressive multiphoton imaging, based on its ability to minimise the space-frequency uncertainty. We implement this approach for the newly developed method of wide-field multiphoton microscopy with single-pixel detection (TRAFIX), which allows imaging through turbid media without correction. The Morlet basis is well-suited to TRAFIX at depth, and promises a route for rapid acquisition with low photodamage.

\end{abstract}

%
\noindent
Optical imaging at depth has gained a strong impetus in the past decade as it allows access to rich and intricate molecular information in three dimensions, even within living animals. 
This is now a burgeoning need in several fields, including neuroscience~\cite{ji_technologies_2016} and  histopathology~\cite{tu_stain-free_2016}.
Researchers are particularly drawn to multiphoton microscopy (MPM), specifically two-photon and latterly three-photon modes, whose near-infrared excitation wavelengths allow deeper penetration into biological tissues~\cite{escobet-montalban_three-photon_2018, wang_three-photon_2018, shemesh_temporally_2017, weisenburger_volumetric_2019}. 
At these depths, to overcome the degradation of beam quality through scattering biological tissues, a range of wavefront shaping methods have been demonstrated~\cite{ji_adaptive_2017}.
However, these methods are challenging, slow, and are typically single-point correction schemes, requiring rapid recalibration when considering any form of wide-field or volumetric imaging.

Rapid MPM may be enabled by the concept of temporal focusing (TF), where the axial localisation is performed by focusing a pulse in time rather than in space~\cite{zhu_simultaneous_2005, oron_scanningless_2005}, alleviating the need for point-scanning with a facile use of a diffracting element.
Recently, TF has come to the forefront with the realisation that spectrally dispersed light preserves spatial fidelity throughout scattering media due to the temporal pulse compression being supported only by the in-phase, minimally scattered photons~\cite{escobet-montalban_wide-field_2018, alemohammad_widefield_2018, wadduwage_-scattering_2019, papagiakoumou_functional_2013}.
Wide-field TF MPM has been demonstrated as a novel option for correction-free imaging at a depth of up to seven scattering mean-free-path lengths with two-photon~\cite{escobet-montalban_wide-field_2018} excitation and may go further using three-photon~\cite{escobet-montalban_wide-field_2019} excitation modes.
This advance (termed TRAFIX) was enabled with single-pixel detection, wherein structured patterns are sequentially projected onto the sample plane.
The total diffuse signal is recorded by a single-pixel detector, and a minimisation algorithm is used to recover the image.
This alleviates the need for spatial coherence in the detection path~\cite{duran_compressive_2015}, enables compressive sensing~\cite{alemohammad_widefield_2018}, and can also be performed in parallel, supporting fast acquisition times~\cite{wadduwage_-scattering_2019}.

Compressive sensing (CS) has led to remarkable achievements~\cite{shapiro_computational_2008, sun_single-pixel_2016}, with a primary advantage of being able to reconstruct images with sampling well below that required by the Nyquist criterion~\cite{candes_robust_2006, donoho_compressed_2006, baraniuk_compressive_2007}.
CS, however, has been applied primarily to macro imaging. 
To date, little consideration has been given to high-resolution microscopy. 
CS may have a number of advantages in this area, including a reduction in photodamage~\cite{woringer_faster_2017}.
CS requires the use of appropriately chosen structured illumination patterns as a measurement basis set. 
Patterns possessing high spatial frequencies are challenging to relay through microscopy systems due to the diffraction limit, resulting in degraded resolution and fidelity and, ultimately, the loss of information about the sample.
In selecting the optimal compression basis, one must consider optimisation of the spatial frequency content, beyond simply using a basis that best suits CS algorithms.

Optimal spatial and frequency sampling has been described by Gabor~\cite{gabor_theory_1946}, presenting the minimal trade-off between spatial and frequency localisation, which reaches the uncertainty limit.
These real-valued Gabor filters (Morlets) have been used to generate a randomised basis for compressive imaging, optimising information transfer to typical frequencies found in photographs~\cite{czajkowski_single-pixel_2018}.
However, this principle is ideally suited for microscopy, where the frequency transfer function is well-known.
In this paper, we examine the random Morlet basis (a convolution of a Gabor filter with a random matrix) as an optimal basis set for CS in TRAFIX.
We demonstrate that the Morlet basis provides superior MPM performance to the conventional Hadamard and Random bases, with CS and when imaging through scattering media.
The elegant formation and optimal performance of the Morlet basis are likely to stimulate its wider adoption for compressive wide-field microscopy. 

%
Pattern efficiency in TRAFIX relies on two principles: the suitability of the basis for CS, and the resilience of the patterned wavefront to propagation and scattering, which we describe in turn. 
CS is achieved by recognising that most signals are close-to sparse in some domain~\cite{baraniuk_compressive_2007}.
Briefly, we consider our linearised image vector $\mathbf{x}$ to comprise a sparse signal $\mathbf{s}$ in some domain, $\Psi$ (here, the discrete cosine transform domain), such that $\mathbf{x}=\Psi \mathbf{s}$.
Given that the majority of images are compressible, it is likely that many coefficients of $\mathbf{s}$ are close to zero~\cite{baraniuk_compressive_2007}.
Thus, images with $N$ total pixels, can be acquired with $M<N$ measurements, where $M$ exceeds the number of non-zero coefficients of $\mathbf{s}$.
We do this by constructing a measurement basis $\Phi$, comprising $M$ rows of linearised patterns with $N$ columns. 
Projecting each row sequentially onto the sample generates measurements on the photodetector given by $\mathbf{y} = \Phi \mathbf{x} = \Phi \Psi \mathbf{s} = \Theta\, \mathbf{s}$.
We estimate the image $\hat{\textbf{x}}$ using $l_1$-norm minimisation: $\hat{\mathbf{x}} = \Psi \hat{\mathbf{s}} = \Psi \cdot \mathrm{arg\,min}||\mathbf{s}||_1, \, s.t. \, \Theta\, \mathbf{s} = \mathbf{y}$, \ie by finding the most sparse $\mathbf{s}$ that can generate the measurements in $\mathbf{y}$.
The efficacy of CS lies in designing $\Phi$ such that any linear combination of columns of $\Theta = \Phi \Psi$ are \textit{mutually incoherent} (\ie posses low correlation between any two columns)~\cite{candes_stable_2006}.
Orthonormal bases, such as the Hadamard, perform well when fully sampled ($M=N$); however, when under-sampled ($M<N$), lead to a $\Theta$ that is not mutually incoherent.
Interestingly, random matrices are mutually incoherent and, as such, can be used for a substantially higher compression~\cite{baraniuk_compressive_2007}.

For high-resolution applications, patterns to be imaged into the sample space are susceptible to degradation arising from spatial filtering in the objective's back focal plane. 
Conventional CS patterns, such as the random pattern, possess very broad spatial frequency spectra.
This leads to the loss of high-frequency components in both illumination and detection, a discrepancy between the generated and the projected $\Psi$, and thus the loss of image quality.
The proposed random Morlet patterns can be contained primarily within the entrance pupil of the objective, thus, they will be transmitted faithfully.

%
The Morlet wavelet is described by a real-valued, centred, zero-mean Gabor filter:
\begin{align}
    g(x,y) = 
    N \cdot 
    & \exp \left\{-\frac{x^2+y^2}{2\sigma^2} \right\}  \times \nonumber \\ 
    & \cos \left\{ \frac{\pi n_p}{2\sigma} \left[x \cos{\theta} + y \sin{\theta} \right] \right\} \, , \label{eq:morlet}
\end{align}
where the first exponential term is a Gaussian with a given space-frequency bandwidth, $\sigma$, and the second term sets a modulation along a given direction $\theta$ that shifts the wavelet in the frequency domain;
$n_p$ is the number of peaks of the Morlet wavelet; and,
$N$ in a normalisation factor, chosen such that $|g|=1$. 

A basis is generated from a set of Morlet wavelets with $\sigma$ and $n_p$ chosen randomly from a normal distribution, convolved with an array of normally distributed random values~\cite{czajkowski_single-pixel_2018}.
The resultant basis, inspired by Gabor's filters~\cite{gabor_theory_1946}, allows for fine spatial features to be sampled, whilst minimising the required spatial frequency bandwidth required.
In particular, it is important to confine the frequency content of the basis to that supported by the imaging system.
For a Morlet pattern illuminating the sample, we can approximate its field in the Fourier plane as the Fourier transform of $g(x,y)$ evaluated at spatial frequencies $(x_2/\lambda f, y_2/\lambda f)$.
A Morlet wavelet in the Fourier plane is described by:
\begin{equation}
    G(x_2,y_2) \propto
    \exp \left\{ 
    -\frac{1}{2\sigma_f^2} \left[ ( x_2 - a_{f}\cos{\theta} )^2+( y_2 - a_{f}\sin{\theta} )^2 \right]
    \right\} \, , \label{eq:morlet_efp_simplified}
\end{equation}
where: $\sigma_f = f \lambda / 2\pi\sigma$ and $a_{f} = f\lambda n_p / 4\sigma$ 
represent the frequency bandwidth and the frequency shift, respectively.
Given a particular back aperture radius, $R = f\cdot \mathrm{NA}$, it is trivial to select $\sigma$ and $n_p$ that fit, for instance, $a_f + 2\sigma_f < R$. 
By contrast, a binary Random basis will uniformly overfill the back aperture to the Nyquist spatial frequency.

\begin{figure}[htbp]
\centering
\includegraphics[width=\linewidth]{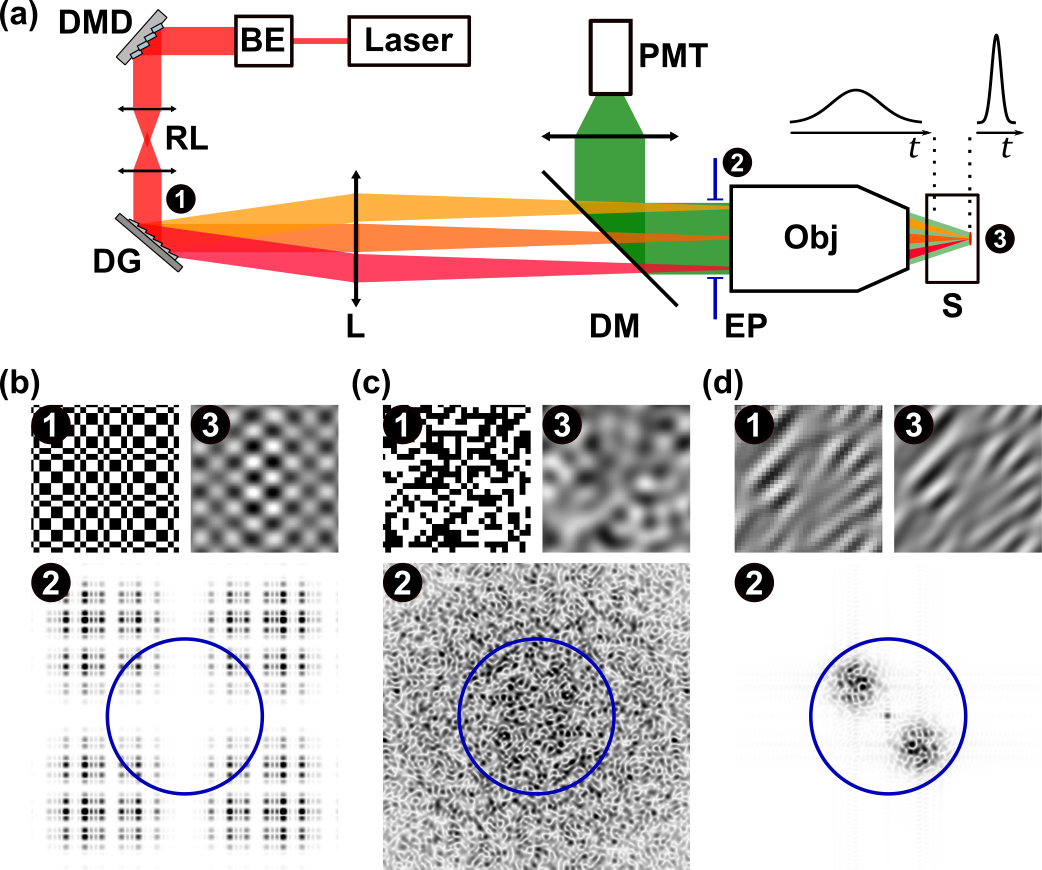}
\caption{Principle of TRAFIX. (a) Optical set-up. BE: beam expander; DMD: digital micromirror device; RL: relay lenses; DG: diffraction grating; L: lens; DM: dichroic mirror; EP: entrance pupil; Obj: objective; S: sample; and, PMT: photomultiplier tube.
Numbered locations correspond to (1) the image on the DG, (2) the Fourier plane of the Obj, and (3) the sample image plane.
(b-d) Simulated field intensity in free space of a Hadamard, Random and Morlet pattern, respectively, at locations (1-3). 
Clipping by the EP is illustrated by a blue circle.}
\label{fig:fig1}
\end{figure}

Fig.~\ref{fig:fig1} illustrates the experimental set up of TRAFIX, where a pulsed laser source (Chameleon Ultra II, Coherent) illuminates a digital micromirror device (DMD; DLP9000, TI). 
The DMD is imaged onto a blazed reflective diffraction grating (DG; 600 lines/mm, Thorlabs), which spatially disperses the light in the back aperture of the objective. 
The dispersed pulse is spatio-temporally refocused in the sample plane by an air immersion objective (20$\times$, 0.75 NA, Nikon).
The resultant multiphoton signal is diffusely collected by a photomultiplier tube (PMT; PMT2101, Thorlabs).
The laser is tuned to a centre wavelength of 800-nm, with a 140-fs pulse duration and an 80-MHz repetition rate.
In-house built MATLAB and C\# software is used to generate the basis, and sequentially project it on the DMD and record the PMT signal.
CS using $l_1$-norm minimisation is performed using the open source `$l_1$-magic' 
toolbox~\citep{candes_robust_2006} and approximated using the `NESTA' algorithm~\cite{becker_nesta:_2011}.

%

%
Fig.~\ref{fig:fig1}(b--d) illustrates representative patterns from Hadamard, Random and Morlet bases, respectively, evaluated using Fourier optics in free space.
An ideal pattern is shaped by the DMD (location (1)), and a Fourier plane (FP) is formed by lens, L, at the back focal plane of the objective (location (2)).
The DC component is omitted for clarity.
The periodic structure of the Hadamard pattern in Fig.~\ref{fig:fig1}(b) leads to a broad, structured, field intensity in the FP, whilst the Random pattern in Fig.~\ref{fig:fig1}(c) leads to a speckle pattern in the FP.
It is evident that both patterns greatly overfill the entrance pupil of the objective, which is marked by the blue circle (15-mm diameter).
This leads to a substantial portion of the field being filtered out by the aperture before reaching the sample.
Thus, the patterns look considerably distorted at the sample image plane (location (3)).
Such distortion is consistent with the qualitative observations in experiment that unless the pattern pixel size is made substantially greater than the diffraction limit, resolution is compromised.
Fig.~\ref{fig:fig1}(c) illustrates that the Morlet field at the FP can be effectively transmitted through the objective, leading to the image plane at the sample and the DG being nearly identical.

\begin{figure}[htbp]
\centering
\includegraphics[width=\linewidth]{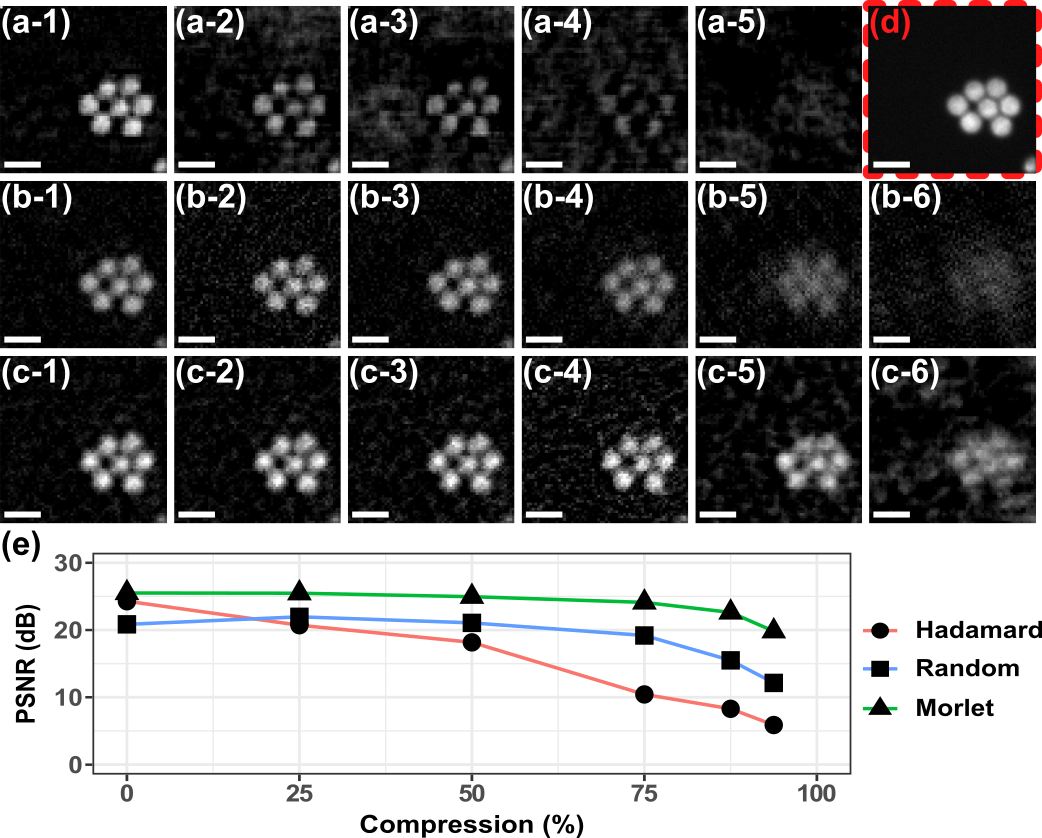}
\caption{TRAFIX of 4.8-$\mu$m beads using 64$\times$64-pixel (a) Hadamard, (b) Random and (c) Morlet bases, with 0, 25, 50, 75, 87.5 and 93.75\% compression marked by labels (1-6), respectively. The red-outlined inset (d) is a reference camera image. The scale bars are 10~$\mu$m. (e) Image quality (PSNR) as a function of compression. }
\label{fig:fig2}
\end{figure}

We experimentally evaluate the bases on compressive imaging of 4.8-$\mu$m green fluorescent polystyrene beads (G0500, Thermo Scientific).
The beads were suspended in water, dried onto a microscope cover slip and sealed using UV-curing optical adhesive (68, Norland), minimising photobleaching to less than 5\% in 2 hours of continuous imaging.
Fig.~\ref{fig:fig2}(a-c) shows the beads imaged with 64$\times$64-pixel bases (4096 patterns) over a 45-$\mu$m field-of-view with various levels of compression, compared to a reference image taken by an EMCCD camera (iXon$^{\text{EM}}$+ 885, Andor) (Fig.~\ref{fig:fig2}(d)).
The illumination intensity and PMT integration time were constant for all recordings; thus, the images are scaled to the same system noise floor.
Image quality is quantified as the peak signal-to-noise ratio (PSNR) (Fig.~\ref{fig:fig2}(e)), defined as: $10 \log_{10}(\textrm{max}(I)^2 / MSE)$, where $I$ is the image intensity and $MSE$ is the mean squared error between the image and the camera reference.
Qualitatively, and from the PSNR, we can see that the Hadamard basis performs well without compression; however, image fidelity declines rapidly with compression.
Even at 25\% compression (\ie using 3/4 of the total patterns), image quality drops more than two fold.
The Random basis performs consistently with compression; however, since it comprises the highest spatial frequency bandwidth of all bases, the maximum achievable PSNR is reduced overall.
The most significant benefit is observed using the Morlet basis, demonstrating the highest PSNR and a high resilience of image quality to compression. 
Remarkably, even at 87.5\% compression, individual beads can be resolved and localised, with a dynamic range above the Hadamard and Random bases at 25\% compression.
It is important to note that since the Random and Morlet bases are not orthonormal, CS recovery at none to low compression leads to an overdetermined measurement matrix whose inverse is ill-conditioned.
To overcome this, the no-compression acquisitions Figs.~\ref{fig:fig2}(b-1) and (c-1) are decomposed into two 50\% compression datasets, which are averaged together.

\begin{figure}[htbp]
\centering
\includegraphics[width=\linewidth]{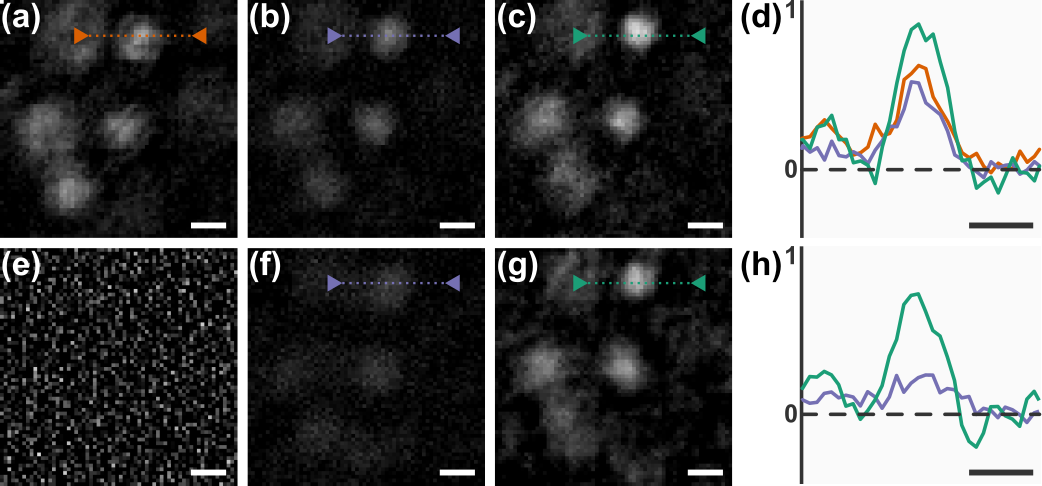}
\caption{TRAFIX images of 4.8-$\mu$m beads through 360~$\mu$m of a scattering phantom, with (a-d) no compression and (e-h) 87.5\% compression; using 64$\times$64-pixel (a,e) Hadamard, (b,f) Random and (c,g) Morlet bases, with corresponding (d,h) line plots from the regions indicated by the colored dashed lines. (e) is omitted in (h) for clarity. The scale bars are 5~$\mu$m.}
\label{fig:fig3}
\end{figure}

We further evaluate the capacity of these bases to image through scattering media.
Fig.~\ref{fig:fig3} shows 4.8-$\mu$m beads imaged through a 360-$\mu$m thick scattering phantom (mean-free-path length, $l_s = 115$~$\mu$m), described in~\cite{escobet-montalban_wide-field_2018}.
At this thickness, multiple scattering of the two-photon signal scrambles spatial information such that no discernible image can be formed at the camera.
However, using single-pixel detection alleviates the need to preserve spatial information, thus, beads can still be resolved (Figs.~\ref{fig:fig3}(a-d)).
Since the phantom is not perfectly flat, not all beads are in the focal plane.
Fig.~\ref{fig:fig3}(d) visualises the intensity across a bead in focus for all bases.
We quantify the image quality as the contrast-to-noise ratio (CNR), calculated as the difference in the mean intensity of the bead and the background, over the standard deviation of the background noise, \ie $CNR = (\bar{I}_{bead}-\bar{I}_{bg})/\sigma_{bg}$.
With no compression, Hadamard, Random and Morlet bases generate an CNR of 6.5, 10 and 13, respectively.
Unlike the PSNR in Fig.~\ref{fig:fig2}(e), the Hadamard performs poorly.
In Fig.~\ref{fig:fig3}(a), we can see that the Hadamard basis generates a non-uniform point-spread funtion (PSF).
The structured, periodic nature of the Hadamard basis may lead to a discrete proportion of the patterns being either transmitted or lost. 
In the image in Fig.~\ref{fig:fig3}(a), this is manifested as larger scale pixelation.
Interestingly, with 87.5\% compression, no discernible image can be reconstructed from the Hadamard, and the CNR becomes 0.2, 4.9 and 9.8, for each respective basis.
It is evident that the Morlet basis generates superior image quality, with and without compression through scattering.
Fig.~\ref{fig:fig3}(h) demonstrates that even at high compression and through scattering, the bead is clearly identified.

\begin{figure}[htbp]
\centering
\includegraphics[width=\linewidth]{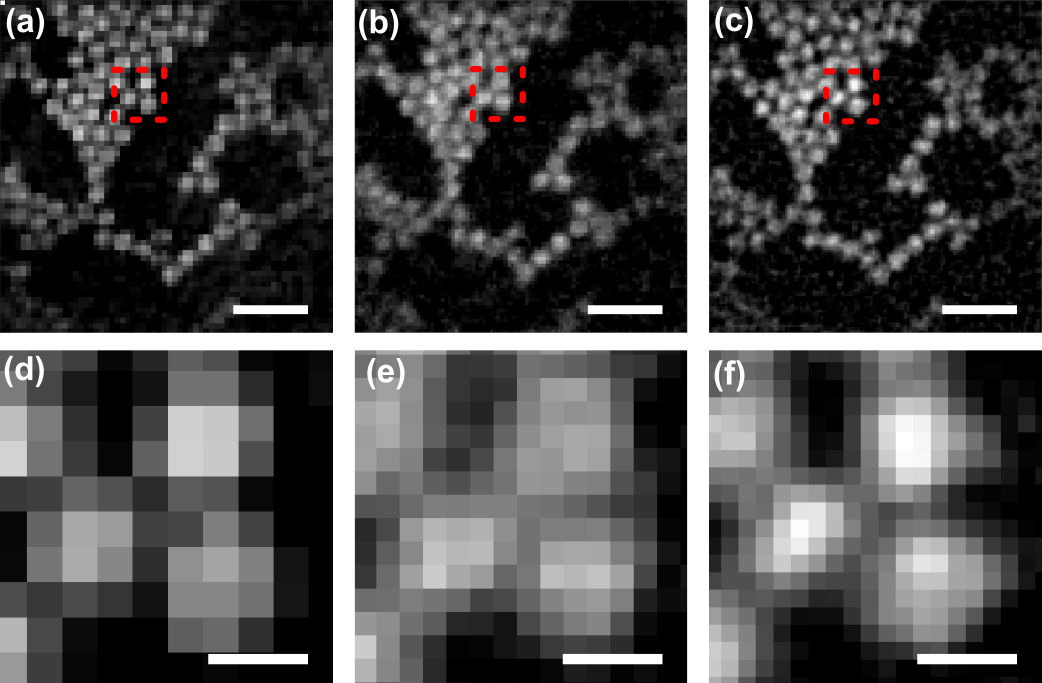}
\caption{TRAFIX images of 10-$\mu$m beads without scattering over a 220-$\mu$m field-of-view using 3072 patterns (equal acquisition time). (a) 64$\times$64-pixel Hadamard, (b) 96$\times$96-pixel Random, (c) 128$\times$128-pixel Morlet bases, corresponding to 25\%, 67\% and 82\% compression. (d,e,f) are the respective magnified insets of (a,b,c), marked by the red-dashed square. The scale bars are 50~$\mu$m for (a,b,c) and 10~$\mu$m for (d,e,f).}
\label{fig:fig4}
\end{figure}

A particular advantage of CS in MPM is that the use of fewer patterns minimises photobleaching, which is an important consideration for sensitive markers, for in vivo and for long-term imaging applications.
If the time of exposure is limited, the Morlet basis should permit an image with the highest sampling resolution for a given image quality.
Fig.~\ref{fig:fig4} demonstrates this by imaging 10-$\mu$m beads without scattering over a 220-$\mu$m field-of-view with a constrained acquisition of 3072 patterns.
Equivalent noise performance is obtained from Hadamard, Random and Morlet bases with 25\%, 67\% and 82\%, respectively, as can be estimated from the PSNR in Fig.~\ref{fig:fig2}(e).
Within these limits, we are able to employ 64-, 96- and 128-pixel wide bases, respectively.
Given the larger sampling, we employ the substantially more efficient NESTA algorithm~\cite{becker_nesta:_2011} that approximately solves the $l_1$-minimisation problem, whilst adhering to a set spatial smoothness, $||\mathbf{y}-\Phi \mathbf{x}||_{l_2} < \epsilon$.
Figs.~\ref{fig:fig4}(d--f) show a close up of the beads.
It is evident that there is a progressive increase in tolerance to higher sampling resolutions, with the Morlet basis clearly delineating beads with the least pixelation.

In this paper, we have demonstrated that the random Morlet basis, based on Gabor's minimisation of the uncertainty criteria, presents an elegant and optimal solution to compressive imaging in microscopy.
Very recently, we have seen a flurry of demonstrations combining wide-field temporal focusing with single-pixel detection to achieve correction-free multiphoton imaging at improved depths~\cite{escobet-montalban_wide-field_2018, alemohammad_widefield_2018, wadduwage_-scattering_2019, escobet-montalban_wide-field_2019}.
Fortuitously, this geometry lends itself to CS, with prospects to increase acquisition speed and minimise photobleaching, which in itself TRAFIX minimises compared to standard point-scanning approaches.
We have demonstrated that unlike many other conventional CS applications, a substantial consideration must be given to the fidelity of the measurement basis as it is relayed through the focusing system and scattered by the sample.

The ideal space-frequency control of the Morlet basis, designed to optimise wavefront propagation through a microscopy system, leads to an overall superior performance in image quality and a high resilience to compression.
Furthermore, the Morlet basis minimises power loss, which is important to MPM, where high illumination power is difficult to achieve over a wide field-of-view with affordable laser sources.
The high compression achievable with the Morlet basis results in a substantial reduction in illumination time, and thus photodamage of the sample~\cite{woringer_faster_2017}.
This is a particular area of concern for two-photon and latterly three-photon microscopy.
In this area, as well as due to its optimal performance, the Morlet basis is well-positioned to make a major impact for compressive wide-field single-pixel multiphoton imaging. 

\paragraph{Acknowledgements} 
We thank the UK Engineering and Physical Sciences Research Council (grant EP/P030017/1) and the European Union's Horizon 2020 Framework Programme (H2020) (675512, BE-OPTICAL). We thank Vanya Metodieva, Wardiya Afshar Saber and Federico M. Gasparoli for helpful discussions and support.

\bibliography{zotero_pw}

\begin{thebibliography}{25}
\providecommand{\natexlab}[1]{#1}
\providecommand{\url}[1]{\texttt{#1}}
\expandafter\ifx\csname urlstyle\endcsname\relax
  \providecommand{\doi}[1]{doi: #1}\else
  \providecommand{\doi}{doi: \begingroup \urlstyle{rm}\Url}\fi

\bibitem[Ji et~al.(2016)Ji, Freeman, and Smith]{ji_technologies_2016}
Na~Ji, Jeremy Freeman, and Spencer~L. Smith.
\newblock Technologies for imaging neural activity in large volumes.
\newblock \emph{Nature Neuroscience}, 19\penalty0 (9):\penalty0 1154--1164,
  September 2016.
\newblock ISSN 1546-1726.
\newblock \doi{10.1038/nn.4358}.
\newblock URL \url{https://www.nature.com/articles/nn.4358}.

\bibitem[Tu et~al.(2016)Tu, Liu, Turchinovich, Marjanovic, Lyngsø, Lægsgaard,
  Chaney, Zhao, You, Wilson, Xu, Dantus, and Boppart]{tu_stain-free_2016}
Haohua Tu, Yuan Liu, Dmitry Turchinovich, Marina Marjanovic, Jens~K. Lyngsø,
  Jesper Lægsgaard, Eric~J. Chaney, Youbo Zhao, Sixian You, William~L. Wilson,
  Bingwei Xu, Marcos Dantus, and Stephen~A. Boppart.
\newblock Stain-free histopathology by programmable supercontinuum pulses.
\newblock \emph{Nature Photonics}, 10\penalty0 (8):\penalty0 534--540, August
  2016.
\newblock ISSN 1749-4893.
\newblock \doi{10.1038/nphoton.2016.94}.
\newblock URL \url{https://www.nature.com/articles/nphoton.2016.94}.

\bibitem[Escobet-Montalbán et~al.(2018{\natexlab{a}})Escobet-Montalbán, Liu,
  Nylk, Gasparoli, Yang, and Dholakia]{escobet-montalban_three-photon_2018}
Adrià Escobet-Montalbán, Pengfei Liu, Jonathan Nylk, Federico~M. Gasparoli,
  Zhengyi Yang, and Kishan Dholakia.
\newblock Three-photon light-sheet fluorescence microscopy.
\newblock \emph{bioRxiv}, page 323790, May 2018{\natexlab{a}}.
\newblock \doi{10.1101/323790}.
\newblock URL \url{https://www.biorxiv.org/content/early/2018/05/16/323790}.

\bibitem[Wang et~al.(2018)Wang, Ouzounov, Wu, Horton, Zhang, Wu, Zhang,
  Schnitzer, and Xu]{wang_three-photon_2018}
Tianyu Wang, Dimitre~G. Ouzounov, Chunyan Wu, Nicholas~G. Horton, Bin Zhang,
  Cheng-Hsun Wu, Yanping Zhang, Mark~J. Schnitzer, and Chris Xu.
\newblock Three-photon imaging of mouse brain structure and function through
  the intact skull.
\newblock \emph{Nature Methods}, 15:\penalty0 789--792, September 2018.
\newblock ISSN 1548-7105.
\newblock \doi{10.1038/s41592-018-0115-y}.
\newblock URL \url{https://www.nature.com/articles/s41592-018-0115-y}.

\bibitem[Shemesh et~al.(2017)Shemesh, Tanese, Zampini, Linghu, Piatkevich,
  Ronzitti, Papagiakoumou, Boyden, and Emiliani]{shemesh_temporally_2017}
Or~A. Shemesh, Dimitrii Tanese, Valeria Zampini, Changyang Linghu, Kiryl
  Piatkevich, Emiliano Ronzitti, Eirini Papagiakoumou, Edward~S. Boyden, and
  Valentina Emiliani.
\newblock Temporally precise single-cell-resolution optogenetics.
\newblock \emph{Nature Neuroscience}, 20\penalty0 (12):\penalty0 1796, December
  2017.
\newblock ISSN 1546-1726.
\newblock \doi{10.1038/s41593-017-0018-8}.
\newblock URL \url{https://www.nature.com/articles/s41593-017-0018-8}.

\bibitem[Weisenburger et~al.(2019)Weisenburger, Tejera, Demas, Chen, Manley,
  Sparks, Traub, Daigle, Zeng, Losonczy, and
  Vaziri]{weisenburger_volumetric_2019}
Siegfried Weisenburger, Frank Tejera, Jeffrey Demas, Brandon Chen, Jason
  Manley, Fraser~T. Sparks, Francisca~Martínez Traub, Tanya Daigle, Hongkui
  Zeng, Attila Losonczy, and Alipasha Vaziri.
\newblock Volumetric {Ca}2+ {Imaging} in the {Mouse} {Brain} {Using} {Hybrid}
  {Multiplexed} {Sculpted} {Light} {Microscopy}.
\newblock \emph{Cell}, 177\penalty0 (4):\penalty0 1050--1066, April 2019.
\newblock ISSN 0092-8674, 1097-4172.
\newblock \doi{10.1016/j.cell.2019.03.011}.
\newblock URL \url{https://www.cell.com/cell/abstract/S0092-8674(19)30273-9}.

\bibitem[Ji(2017)]{ji_adaptive_2017}
Na~Ji.
\newblock Adaptive optical fluorescence microscopy.
\newblock \emph{Nature Methods}, 14\penalty0 (4):\penalty0 374--380, April
  2017.
\newblock ISSN 1548-7105.
\newblock \doi{10.1038/nmeth.4218}.
\newblock URL \url{https://www.nature.com/articles/nmeth.4218}.

\bibitem[Zhu et~al.(2005)Zhu, Howe, Durst, Zipfel, and
  Xu]{zhu_simultaneous_2005}
Guanghao Zhu, James~van Howe, Michael Durst, Warren Zipfel, and Chris Xu.
\newblock Simultaneous spatial and temporal focusing of femtosecond pulses.
\newblock \emph{Optics Express}, 13\penalty0 (6):\penalty0 2153--2159, March
  2005.
\newblock ISSN 1094-4087.
\newblock \doi{10.1364/OPEX.13.002153}.
\newblock URL
  \url{https://www.osapublishing.org/oe/abstract.cfm?uri=oe-13-6-2153}.

\bibitem[Oron et~al.(2005)Oron, Tal, and Silberberg]{oron_scanningless_2005}
Dan Oron, Eran Tal, and Yaron Silberberg.
\newblock Scanningless depth-resolved microscopy.
\newblock \emph{Optics Express}, 13\penalty0 (5):\penalty0 1468--1476, March
  2005.
\newblock ISSN 1094-4087.
\newblock \doi{10.1364/OPEX.13.001468}.
\newblock URL
  \url{https://www.osapublishing.org/oe/abstract.cfm?uri=oe-13-5-1468}.

\bibitem[Escobet-Montalbán et~al.(2018{\natexlab{b}})Escobet-Montalbán,
  Spesyvtsev, Chen, Saber, Andrews, Herrington, Mazilu, and
  Dholakia]{escobet-montalban_wide-field_2018}
Adrià Escobet-Montalbán, Roman Spesyvtsev, Mingzhou Chen, Wardiya~Afshar
  Saber, Melissa Andrews, C.~Simon Herrington, Michael Mazilu, and Kishan
  Dholakia.
\newblock Wide-field multiphoton imaging through scattering media without
  correction.
\newblock \emph{Science Advances}, 4\penalty0 (10):\penalty0 eaau1338, October
  2018{\natexlab{b}}.
\newblock ISSN 2375-2548.
\newblock \doi{10.1126/sciadv.aau1338}.
\newblock URL \url{http://advances.sciencemag.org/content/4/10/eaau1338}.

\bibitem[Alemohammad et~al.(2018)Alemohammad, Shin, Tran, Stroud, Chin, Tran,
  and Foster]{alemohammad_widefield_2018}
Milad Alemohammad, Jaewook Shin, Dung~N. Tran, Jasper~R. Stroud, Sang~Peter
  Chin, Trac~D. Tran, and Mark~A. Foster.
\newblock Widefield compressive multiphoton microscopy.
\newblock \emph{Optics Letters}, 43\penalty0 (12):\penalty0 2989--2992, June
  2018.
\newblock ISSN 1539-4794.
\newblock \doi{10.1364/OL.43.002989}.
\newblock URL
  \url{https://www.osapublishing.org/ol/abstract.cfm?uri=ol-43-12-2989}.

\bibitem[Wadduwage et~al.(2019)Wadduwage, Park, Boivin, Xue, and
  So]{wadduwage_-scattering_2019}
Dushan~N. Wadduwage, Jong~Kang Park, Josiah~R. Boivin, Yi~Xue, and Peter T.~C.
  So.
\newblock De-scattering with {Excitation} {Patterning} ({DEEP}) {Enables}
  {Rapid} {Wide}-field {Imaging} {Through} {Scattering} {Media}.
\newblock \emph{arXiv:1902.10737 [physics]}, February 2019.
\newblock URL \url{http://arxiv.org/abs/1902.10737}.

\bibitem[Papagiakoumou et~al.(2013)Papagiakoumou, Bègue, Leshem, Schwartz,
  Stell, Bradley, Oron, and Emiliani]{papagiakoumou_functional_2013}
Eirini Papagiakoumou, Aurélien Bègue, Ben Leshem, Osip Schwartz, Brandon~M.
  Stell, Jonathan Bradley, Dan Oron, and Valentina Emiliani.
\newblock Functional patterned multiphoton excitation deep inside scattering
  tissue.
\newblock \emph{Nature Photonics}, 7\penalty0 (4):\penalty0 274--278, April
  2013.
\newblock ISSN 1749-4893.
\newblock \doi{10.1038/nphoton.2013.9}.
\newblock URL \url{https://www.nature.com/articles/nphoton.2013.9}.

\bibitem[Escobet-Montalbán et~al.(2019)Escobet-Montalbán, Wijesinghe, Chen,
  and Dholakia]{escobet-montalban_wide-field_2019}
Adrià Escobet-Montalbán, Philip Wijesinghe, Mingzhou Chen, and Kishan
  Dholakia.
\newblock Wide-field multiphoton imaging with {TRAFIX}.
\newblock In \emph{Multiphoton {Microscopy} in the {Biomedical} {Sciences}
  {XIX}}, volume 10882, page 108821G. SPIE BiOS, February 2019.
\newblock \doi{10.1117/12.2508373}.
\newblock URL
  \url{https://www.spiedigitallibrary.org/conference-proceedings-of-spie/10882/108821G/Wide-field-multiphoton-imaging-with-TRAFIX/10.1117/12.2508373.short}.

\bibitem[Durán et~al.(2015)Durán, Soldevila, Irles, Clemente, Tajahuerce,
  Andrés, and Lancis]{duran_compressive_2015}
V.~Durán, F.~Soldevila, E.~Irles, P.~Clemente, E.~Tajahuerce, P.~Andrés, and
  J.~Lancis.
\newblock Compressive imaging in scattering media.
\newblock \emph{Optics Express}, 23\penalty0 (11):\penalty0 14424--14433, June
  2015.
\newblock ISSN 1094-4087.
\newblock \doi{10.1364/OE.23.014424}.
\newblock URL
  \url{https://www.osapublishing.org/oe/abstract.cfm?uri=oe-23-11-14424}.

\bibitem[Shapiro(2008)]{shapiro_computational_2008}
Jeffrey~H. Shapiro.
\newblock Computational ghost imaging.
\newblock \emph{Physical Review A}, 78\penalty0 (6):\penalty0 061802, December
  2008.
\newblock \doi{10.1103/PhysRevA.78.061802}.
\newblock URL \url{https://link.aps.org/doi/10.1103/PhysRevA.78.061802}.

\bibitem[Sun et~al.(2016)Sun, Edgar, Gibson, Sun, Radwell, Lamb, and
  Padgett]{sun_single-pixel_2016}
Ming-Jie Sun, Matthew~P. Edgar, Graham~M. Gibson, Baoqing Sun, Neal Radwell,
  Robert Lamb, and Miles~J. Padgett.
\newblock Single-pixel three-dimensional imaging with time-based depth
  resolution.
\newblock \emph{Nature Communications}, 7:\penalty0 12010, July 2016.
\newblock ISSN 2041-1723.
\newblock \doi{10.1038/ncomms12010}.
\newblock URL \url{https://www.nature.com/articles/ncomms12010}.

\bibitem[Candes et~al.(2006{\natexlab{a}})Candes, Romberg, and
  Tao]{candes_robust_2006}
E.~J. Candes, J.~Romberg, and T.~Tao.
\newblock Robust uncertainty principles: exact signal reconstruction from
  highly incomplete frequency information.
\newblock \emph{IEEE Transactions on Information Theory}, 52\penalty0
  (2):\penalty0 489--509, February 2006{\natexlab{a}}.
\newblock ISSN 0018-9448.
\newblock \doi{10.1109/TIT.2005.862083}.

\bibitem[Donoho(2006)]{donoho_compressed_2006}
D.~L. Donoho.
\newblock Compressed {Sensing}.
\newblock \emph{IEEE Trans. Inf. Theor.}, 52\penalty0 (4):\penalty0 1289--1306,
  April 2006.
\newblock ISSN 0018-9448.
\newblock \doi{10.1109/TIT.2006.871582}.
\newblock URL \url{https://doi.org/10.1109/TIT.2006.871582}.

\bibitem[Baraniuk(2007)]{baraniuk_compressive_2007}
R.~G. Baraniuk.
\newblock Compressive {Sensing} [{Lecture} {Notes}].
\newblock \emph{IEEE Signal Processing Magazine}, 24\penalty0 (4):\penalty0
  118--121, July 2007.
\newblock ISSN 1053-5888.
\newblock \doi{10.1109/MSP.2007.4286571}.

\bibitem[Woringer et~al.(2017)Woringer, Darzacq, Zimmer, and
  Mir]{woringer_faster_2017}
Maxime Woringer, Xavier Darzacq, Christophe Zimmer, and Mustafa Mir.
\newblock Faster and less phototoxic 3d fluorescence microscopy using a
  versatile compressed sensing scheme.
\newblock \emph{Optics Express}, 25\penalty0 (12):\penalty0 13668--13683, June
  2017.
\newblock ISSN 1094-4087.
\newblock \doi{10.1364/OE.25.013668}.
\newblock URL
  \url{https://www.osapublishing.org/oe/abstract.cfm?uri=oe-25-12-13668}.

\bibitem[Gabor(1946)]{gabor_theory_1946}
D.~Gabor.
\newblock Theory of communication. {Part} 1: {The} analysis of information.
\newblock \emph{Journal of the Institution of Electrical Engineers - Part III:
  Radio and Communication Engineering}, 93\penalty0 (26):\penalty0 429--441,
  November 1946.
\newblock ISSN 2054-0604.
\newblock \doi{10.1049/ji-3-2.1946.0074}.
\newblock URL
  \url{https://digital-library.theiet.org/content/journals/10.1049/ji-3-2.1946.0074}.

\bibitem[Czajkowski et~al.(2018)Czajkowski, Pastuszczak, and
  Kotyński]{czajkowski_single-pixel_2018}
Krzysztof~M. Czajkowski, Anna Pastuszczak, and Rafał Kotyński.
\newblock Single-pixel imaging with {Morlet} wavelet correlated random
  patterns.
\newblock \emph{Scientific Reports}, 8\penalty0 (1):\penalty0 466, January
  2018.
\newblock ISSN 2045-2322.
\newblock \doi{10.1038/s41598-017-18968-6}.
\newblock URL \url{https://www.nature.com/articles/s41598-017-18968-6}.

\bibitem[Candes et~al.(2006{\natexlab{b}})Candes, Romberg, and
  Tao]{candes_stable_2006}
E.~J. Candes, J.~Romberg, and T.~Tao.
\newblock Stable signal recovery from incomplete and inaccurate measurements.
\newblock \emph{Communications on Pure and Applied Mathematics}, 59\penalty0
  (8):\penalty0 1207--1223, August 2006{\natexlab{b}}.
\newblock ISSN 0010-3640.
\newblock \doi{10.1002/cpa.20124}.
\newblock URL \url{https://onlinelibrary.wiley.com/doi/abs/10.1002/cpa.20124}.

\bibitem[Becker et~al.(2011)Becker, Bobin, and Candès]{becker_nesta:_2011}
S.~Becker, J.~Bobin, and E.~Candès.
\newblock {NESTA}: {A} {Fast} and {Accurate} {First}-{Order} {Method} for
  {Sparse} {Recovery}.
\newblock \emph{SIAM Journal on Imaging Sciences}, 4\penalty0 (1):\penalty0
  1--39, January 2011.
\newblock \doi{10.1137/090756855}.
\newblock URL \url{https://epubs.siam.org/doi/abs/10.1137/090756855}.

\end{thebibliography}


\end{document}